# Computationally Efficient CFD Prediction of Bubbly Flow using Physics-Guided Deep Learning

Han Bao[1], Jinyong Feng[2*], Nam Dinh[3], Hongbin Zhang[1]

[1] *Idaho National Laboratory, P.O. Box 1625, MS 3860, Idaho Falls, 83415, ID, USA*
[2] *Massachusetts Institute of Technology, Cambridge, 02139, Massachusetts, USA*
[3] *North Carolina State University, Raleigh, 27695, North Carolina, USA*

Abstract:

To realize efficient computational fluid dynamics (CFD) prediction of two-phase flow, a multi-scale framework was proposed in this paper by applying a physics-guided data-driven approach. Instrumental to this framework, Feature Similarity Measurement (FSM) technique was developed for error estimation in two-phase flow simulation using coarse-mesh CFD, to achieve a comparable accuracy as fine-mesh simulations with fast-running feature. By defining physics-guided parameters and variable gradients as physical features, FSM has the capability to capture the underlying local patterns in the coarse-mesh CFD simulation. Massive low-fidelity data and respective high-fidelity data are used to explore the underlying information relevant to the main simulation errors and the effects of phenomenological scaling. By learning from previous simulation data, a surrogate model using deep feedforward neural network (DFNN) can be developed and trained to estimate the simulation error of coarse-mesh CFD. In a demonstration case of two-phase bubbly flow, the DFNN model well captured and corrected the unphysical "peaks" in the velocity and void fraction profiles near the wall in the coarse-mesh configuration, even for extrapolative predictions. The research documented supports the feasibility of the physics-guided deep learning methods for coarse mesh CFD simulations which has a potential for the efficient industrial design.

Keywords: deep learning, two-phase bubbly flow, coarse-mesh CFD, physical feature, data similarity

---

[*] Corresponding author. Tel,: +1 919 645 8225
*E-mail address*: han.bao@inl.gov; fengjinyong2008@gmail.com; ntdinh@ncsu.edu; hongbin.zhang@inl.gov.

# 1. Introduction

Owing to the advancement of high-performance computing and computational methods, modeling and numerical simulations have become instrumental in the design, analysis and licensing of nuclear power plants. Compared to system codes using lumped-parameter models, computational fluid dynamics (CFD) methods have been widely used for solving transport equations of fluid mechanics by using local instantaneous formulations with finer mesh sizes, where small-scale flow features could be captured. While CFD has the potential to accurately predict the flow behavior and reduce the need for dedicated reactor-operational experiments, it suffers from three key challenges for the system-level analysis of NPP behaviors, namely high computational costs, user effects, and limited understanding on error sources of CFD simulation.

The main limitation of applying CFD methods to practical industrial applications is the computational cost. Since discretizing the temporal and spatial space on a much smaller scale, CFD simulations require many more cells than a system thermal-hydraulic simulation. One of the most representative examples is direct numerical simulation (DNS) method. As a first principle based method, DNS directly solves the Navier-Stokes equations without any closure models, thus making it serve as high fidelity benchmark data, especially in the single-phase study. By coupling with interface tracking method (Hirt and Nichols, 1981; Sussman et al., 1994; Unverdi and Tryggvason, 1992), DNS extends its capability to simulate two-phase flow. In the state-of-the-art two-phase DNS simulation (Fang et al., 2018), in order to resolve each individual bubble and turbulent eddies down to the smallest turbulent length scale, i.e., Kolmogorov scale (Kolmogorov, 1941), it requires 1.10 billion cells and ~730,000 core-hours to simulate the reactor subchannel with hydraulics Reynolds number of 80,000 whereas the hydraulics Reynolds number under reactor operational conditions is ~500,000. Productive CFD simulations have to be performed on large supercomputers rather than a multi-core computer. To bypass the computational cost of the fully resolved high Reynolds number case, researchers either conducted separate effect studies with well-controlled flow conditions (Bunner and Tryggvason, 2003; Feng and Bolotnov, 2017a, 2017b; Feng and Bolotnov, 2017) to develop individual closures, or adopt computational efficient Reynolds-averaging Navier-Stokes method (Brewster et al., 2015; Feng et al., 2018).

Another aspect leading to the limitation of CFD simulation on system-level analysis is the user effect, particularly on multiphase flow CFD. CFD codes are designed to be general flow solvers, applicable to nearly every scale of flow problem encountered in engineering practice, spanning from high Mach number compressible air flow to flow through rod bundles in nuclear reactors. Even with the same CFD code and same specifications (e.g. initial and boundary conditions), users' knowledge is not always complete but relies on the selection of closure models and mesh sizes. The small number of modeled physical processes have a few of associated models that work well for some scale of problems but not well for others. For example, the Lemmert-Chawla's (Lemmert and Chawla, 1977) nucleation site density model predicts acceptable nucleation site at atmospheric pressure. However, it is not applicable at high pressure due to the lack of pressure dependence (Gilman and Baglietto, 2017). Currently, the mesh size and models are selected based on previous modeling experience, this kind of "educated guess" may lead to errors in treatment of new physical conditions.

The last limitation is the limited understanding on error sources of CFD simulations. There are two major error sources in the CFD applications: physical model error and mesh-induced numerical error. Physical model error arises from physical assumptions and mathematical approximations in the form of the model equations (misrepresentation of the underlying physical system) and errors in assigning values for model parameters or calibration coefficients. Mesh-induced numerical error comes from the solution discretization in space and time, and the approximation used in over-cell integration of the variables that are non-uniformly distributed within the cell. While CFD simulations are advantageous for its high fidelity and resolution, these benefits have largely not been realized in reactor safety applications due to the lack of established and agreed upon approaches for assessing those error sources in CFD simulations. Two mainstream research directions of error estimation in CFD simulations are numerical error estimation, which are translated as the mesh error or discretization scheme error (Eça and Hoekstra, 2014; Ferziger and Peric, 2012), and model error estimation where the model error estimations are usually performed by varying the model coefficients (Edeling et al., 2014; Kato and Obayashi, 2013; Liu et al., 2019a, 2019b; Liu and Dinh, 2019; Wu et al., 2018a, 2018b) to justify the solution range. In two-phase flow CFD simulation, two main error sources are both tightly connected with local mesh sizes, which makes it difficult to analyze them separately. Furthermore, some of the applied models in CFD codes are not scalable for extrapolative predictions because of the lack of validation data during the development phase. The uncertainty stemming from using these models increases outside of their applicable ranges. Mesh configuration is crucial to the accurate prediction of the multiphase flow phenomena using CFD approaches. The calculation of flow variables gradients depends on the mesh resolution between two adjacent cells which are directly relevant to certain physical models. For example, the lift force relies on the gradient of velocity and the magnitude of turbulence dispersion force depends on the gradient of void fraction. In addition, the near wall mesh resolution determines numerous closures near the wall, like velocity wall function. The lift coefficient model proposed by Shaver and Podowski (Shaver and Podowski, 2015) also depends on the ratio between wall distance and interaction length scale.

To deal with these difficulties, this paper applied a data-driven approach, Feature Similarity Measurement (FSM), for error estimation in two-phase flow simulation using coarse-mesh CFD, to achieve a comparable accuracy as fine-mesh simulations with the featured fast-running capability. The approach was preliminarily developed by the authors before where results demonstrated good predictions on single phase flow (Bao et al., 2018) and mixed convection (Bao et al., 2019b, 2019a) using GOTHIC as low-fidelity simulation tool. As a coarse-mesh thermal-hydraulic simulation tool, GOTHIC has been widely used for reactor safety analysis (Bao et al., 2018). The performance of this proposed approach is also evaluated considering the interpolative and extrapolative predictions. Computational cost for system-level thermal hydraulic modeling and simulation could be reduced by using coarse-mesh CFD, meanwhile, this data-driven approach treats model error and mesh-induced numerical error together by taking their tight connection with local mesh size into consideration.

In this paper, the data-driven approach is further developed and demonstrated on a two-phase bubbly flow case study. Comprehensive description of the proposed approach is provided in Section 2 which provides a guidance about how to use deep learning to explore local physics and bridge the global scale gap. Section 3 illustrates the workflow of applying FSM to realize computationally efficient CFD simulation, which is demonstrated using a bubbly flow case study in Section 4 and further discussed in Section 5. Key findings and future works are summarized in Section 6.

## 2. Technical background

*2.1. Recent machine learning applications in CFD thermal-hydraulic simulation*

Among all CFD methods, RANS (Reynolds-averaged Navier-Stokes) approach remains the standard in industry applications since it can model flows in complex geometries involving coupled phenomena in different timescales compatible with industrial constraints. RANS approach represents the entire spectrum of the flow turbulence with empirical turbulence models, which rely on strong simplifying assumptions that are not applicable for realistic applications. This approach has limited accuracy, especially for two-phase flow simulation, while industry expects an increasing predictive performance for extrapolations. Recently, the rise of performance computing has led to a large high-fidelity data generation from DNS and well-resolved LES (large eddy simulation) for the training and development of data-driven turbulence closures. Different machine learning techniques, e.g. neural networks (NNs), Random Forests (RFs), have been widely used to predict different relevant parameters or source terms for turbulence transport equations (Hanna, 2018; Ling et al., 2016; Tracey et al., 2015; Wang et al., 2017). A major part of these machine learning applications focused on single phase flow and how to improve RANS turbulence modeling without considering mesh-induced numerical errors. Turbulence model error is estimated in different validation domains, but high resolution still results to high computational cost. Similarly, a data-driven approach was developed to estimate the model error from boiling closures (Liu et al., 2018). Hanna (Hanna, 2018) proposed a coarse grid CFD approach using machine learning algorithms to predict the local errors. This work aims at the correction of mesh-induced numerical error without considering the model errors that may affect results in thermal hydraulic analysis. These efforts analyzed model error and mesh-induced numerical error separately with another fixed, which is inapplicable for two-phase flow where mesh size is treated as a key model parameter and fine-mesh convergence is not expected. The uncertainty propagation due to scaling issues makes it more difficult to estimate the simulation error when using these codes for realistic system-level NPP analysis. The proposed data-driven approach, FSM (Bao et al., 2019a), integrates model error and mesh-induced numerical error together. The deep learning model trained in FSM approach treats the physical correlations, coarse mesh sizes and numerical solvers as an integrated model, which can be considered as a surrogate of governing equations and closure correlations of coarse-mesh CFD.

*2.2. Classification of physics coverage condition: a potential to enlarge the validation domain*

There are two "ideal" approaches to explore and predict behaviors in prototype-scale applications: (1) prototype-scale experiment, which presumably duplicates a prototype-scale phenomenon existing in the real applications, (2) DNS modeling where the local information is solved accurately with very fine mesh. However, prototype-scale experiments are hard to build while many prototype-scale tests are required in the real applications, and DNS is computationally expensive to deal with the system-level predictions. Traditional CFD approaches, such as RANS, are treated as reduced-order models because their predictions of prototype-scale processes are made using models developed based on scaled experiments. However, these reduced-order CFD approaches are still not practical for wide use in system-level NPP analysis because of high computational cost compared with systems codes. The development of FSM enables CFD approaches with capability to realize the computationally efficient prediction by exploring local physical features instead of global characteristics.

Over the past few decades, many concepts of nuclear reactor have been proposed with different components, geometries, and powers. The respective global characteristics might be out of the domain of previous designs or simulations, which brings large uncertainty into the application. The relevant thermal-hydraulic experiments with a wide range of scale and structure must be designed and implemented for code validation and licensing of new reactor designs. The extrapolation of global characteristics, e.g., dimension, geometry/structure, initial conditions/boundary conditions (ICs/BCs), or powers, may limit the applicability of the previously developed models or experiments. However, local physics such as the interaction between liquid, vapor and heat structure may not change significantly: quantities of interest (temperature, velocities and vapor fraction) remain approximately the same although their different values lead to different flow patterns (Bao et al., 2019a). This makes it possible that some well-defined local physical features in the local cells are similar even if the global characteristics vary significantly.

According to the similarity or coverage of global characteristics and local physical features, four different physics coverage conditions (PCCs) are classified: global interpolation through local interpolation (GILI), global interpolation through local extrapolation (GILE), global extrapolation through local interpolation (GELI) and global extrapolation through local extrapolation (GELE) (Bao et al., 2019b). The GELI condition refers to the situation where the global characteristics of target case are identified as an extrapolation of existing cases, but the local physical features of target case are similar to or mostly covered by the ones of existing cases. These well-defined local physical features are supposed to represent the specific underlying local physics. The interpolation or similarity of local physics between the target case and existing cases depends on the identification of physical features, data quality and quantity. The local similarity in GELI condition makes it feasible to derive benefits from the existing data to estimate the target case. Instead of endlessly evaluating the applicable ranges of models and scaling uncertainty in extrapolative predictions, exploring the similarity of local physics provides a potential to bridge the scale gap in global extrapolations.

Although some advanced codes have been widely used in system-level NPP analysis, the V&V of these codes still suffer from the lack of validation data. The application domain for new rector designs are always not met by the validation domain defined based on global characteristics, as shown in the left part of Figure 1. A major fraction of the validation domain belongs to GILI condition, the grounded physics coverage condition for code/model V&V, where the existing data or models has the capability to estimate the target case due to both the global and local similarities. For the GILE condition, even if global physical condition of the target case is covered by existing cases, data from existing cases is not able to predict the target case since local physics are different. For instance, the models developed from experiments of laminar flow or turbulent flow are not applicable for transition prediction, although the global Reynolds number is covered. In contrast, the GELI condition has a potential to be added into the validation domain once the similarity of local physical features can be well defined and determined. From the perspective of data analysis, the previous validation domain defined by global characteristics may be expanded if it is separated into several new validation domains re-classified by local physical features, as illustrated in the right part of Figure 1. Focusing on GELI condition, FSM has a potential to provide insights on the designs of experiments and numerical tests to enlarge the validation domain to reach the required application domain.

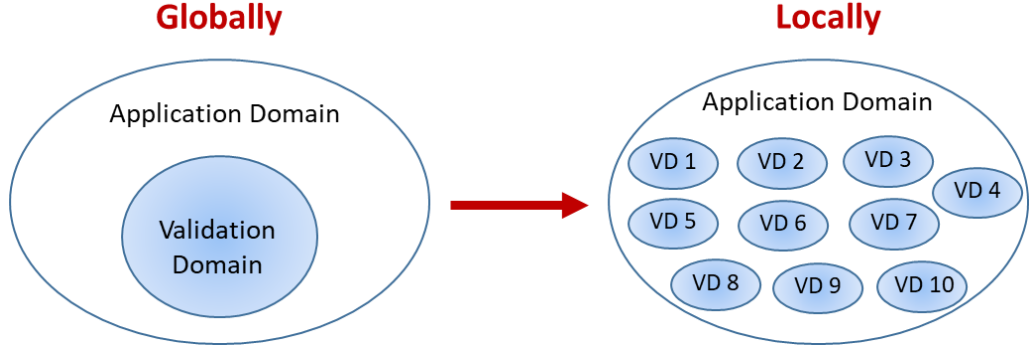

Figure 1. Exploring similarity of local physical features: a way to re-classify and enlarge the validation domain to reach the application domain.

*2.3. Deep learning technique: deep feedforward neural network (DFNN)*

The application of deep feedforward neural network (DFNN) in this work is to fit the function between simulation error and several physical feature inputs. A DFNN normally includes several hidden layers with a couple of neurons and activation functions on each hidden layer. By using multiple layers of transformations, deep neural networks are able to capture complex, hierarchical interactions between features.

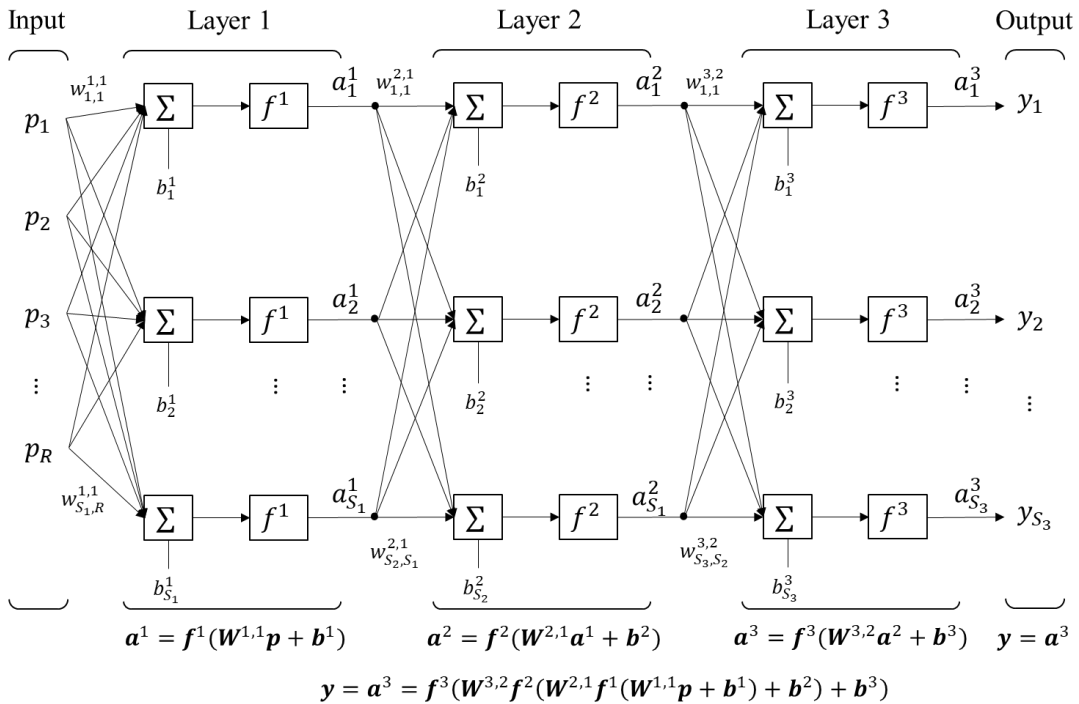

Figure 2. A typical three-hidden-layer feedforward network.

Figure 2 illustrated the schematic of a typical three-hidden-layer feedforward network with one input layer, three hidden layers, and one output layer. The network shown above has $\boldsymbol{p}_{R*1}$ inputs, $S_1$ neurons in the first layer, $S_2$ and $S_3$ neurons in the second and third layer, etc. Each layer (layer $i$) has a weight matrix $\boldsymbol{W}_{S_i*S_{i-1}}$, a bias vector $\boldsymbol{b}_{S_i*1}$, and an output vector $\boldsymbol{a}_{S_i*1}$. $\boldsymbol{W}_{S_i*S_{i-1}}$ and $\boldsymbol{b}_{S_i*1}$ are both adjustable scalar parameters of the neuron. Here the output of the third layer ($\boldsymbol{a}^3$) is the network output of interest, labeled as $\boldsymbol{y}$, which is a function of inputs $\boldsymbol{p}$ fitted by a DFNN model. The information flow is straightforward from input to output, so it is called feedforward network. The three-hidden-layer DFNN includes the following (non-)linear transformations:

$$\boldsymbol{a}^1 = \boldsymbol{f}^1(\boldsymbol{W}^{1,1}\boldsymbol{p} + \boldsymbol{b}^1) \tag{1}$$

$$\boldsymbol{a}^2 = \boldsymbol{f}^2(\boldsymbol{W}^{2,1}\boldsymbol{a}^1 + \boldsymbol{b}^2) \tag{2}$$

$$\boldsymbol{a}^3 = \boldsymbol{f}^3(\boldsymbol{W}^{3,2}\boldsymbol{a}^2 + \boldsymbol{b}^3) \tag{3}$$

$$\boldsymbol{y} = \boldsymbol{a}^3 = \boldsymbol{f}^3(\boldsymbol{W}^{3,2}\boldsymbol{f}^2(\boldsymbol{W}^{2,1}\boldsymbol{f}^1(\boldsymbol{W}^{1,1}\boldsymbol{p} + \boldsymbol{b}^1) + \boldsymbol{b}^2) + \boldsymbol{b}^3) \tag{4}$$

Here $f^i$ is an activation function, typically a step function, a rectified linear unit, a tan-sigmoid function or a log-sigmoid function, which produces the output $\boldsymbol{a}^i$. The tanh-sigmoid activation function is commonly used in backpropagation networks because it is differentiable and non-linear. The tanh-sigmoid activation function is applied in this work, as expressed below,

$$f(x) = \frac{e^x - e^{-x}}{e^x + e^{-x}} \tag{5}$$

Once the network weights and biases have been initialized, the network is ready for training. The training process requires a set of examples of proper network behavior including inputs $\boldsymbol{p}$ and target outputs $\boldsymbol{t}$. During training the weights and biases of the network are iteratively adjusted to minimize the network prediction error. The evaluation metric of the function fitting is the Mean Squared Error (MSE) at each test point,

$$mse = \frac{1}{N_d} \sum_{k=1}^{N_d} (\boldsymbol{t}_k - \boldsymbol{y}_k)^2 \tag{6}$$

where $N_d$ is the number of training data points. The central idea of neural networks is that such parameters (weights and biases) can be adjusted so that the network exhibits some desired nonlinear or comprehensive behaviors. There are many algorithms to adjust the weights and biases using the gradient of cost function (error) to determine how to adjust the weights to minimize the error. The gradient is determined using a technique called backpropagation, which involves error computations backwards through the network by generalizing the gradient descendent rule to multiple-layer networks and nonlinear differentiable activation functions. Training input vectors and the corresponding target vectors are used to train a network until it can approximate a function that associates input vectors with specific output vectors. Then testing data set is applied for selecting the optimum number of iterations to avoid overfitting. Once the learning process ends, another data set (validation data set) is used to validate and confirm the prediction accuracy for new inputs. Levenberg-Marquardt algorithm (Hagan and Menhaj, 1994) is recommended for small or medium problems and usually runs fast, but the drawback is that it requires to store some matrices which can be quite large for certain problems. As the number of weights increases, its advantage of fast convergence decreases. Another algorithm that often provides fast convergence, batch gradient descent with momentum (Rehman and Nawi, 2011), makes weight changes equal to the sum of a fraction of the last weight change and the new change suggested by the backpropagation rule. It allows greater learning rates while maintaining stability, but still too slow for many practical applications. For some noisy and challenging problems, Bayesian regularization (MacKay, 1992) takes longer time but obtains a better solution. This algorithm can well prevent the problem that the network can overfit on the training set and not generalize well to new data outside the training set. The typical cost function used for training an FNN is the MSE as shown in Equation (**6**). By adding a term that consists of the mean of the sum of squares of the weights and biases, the cost function can be modified to improve the generalization (Demuth and Beale, 1998):

$$msereg = \gamma \cdot mse + (1 - \gamma) msw \tag{7}$$

$$msw = \frac{1}{N_w} \sum_{j=1}^{N_w} w_j^2 \tag{8}$$

$$S(\boldsymbol{W}) = \gamma \cdot \frac{1}{N_d}\sum_{k=1}^{N_d}(\boldsymbol{t}_k - \boldsymbol{y}_k)^2 + (1-\gamma)\frac{1}{N_w}\sum_{j=1}^{N_w} w_j^2 \qquad (9)$$

where $\gamma \in [0,1]$ is the regularization parameter, $N_w$ is the number of weights, for this 3-hidden-layer DFNN, $m = S_1 R + S_2 S_1 + S_3 S_2$. The new cost function for hidden layer 3, $S(\boldsymbol{W})$, will help reduce the number of weights and biases to be smaller than the number of training data points, which will force the network response to be smoother and less likely to overfit. The difficulty of performing regularization is how to determine the optimum value for the performance ratio parameter. In the Bayesian framework of MacKay (MacKay, 1992), the values of optimal regularization parameters are estimated via statistical techniques. Weights and biases of the network are assumed to be random variables within some specified distributions. Bayesian regularization is used as the training function to adjust the weights and biases in this work. More details about Bayesian regularization algorithm for backward propagation derived by MacKay (MacKay, 1992) and Burden (Burden and Winkler, 2008) can be found in references.

## 3. Proposed data-driven approach for computationally efficient CFD simulation

The proposed data-driven approach, Feature Similarity Measurement (FSM), was developed to identify the local physical features, measure the data similarity of defined physical features, and investigate the relationship between physical feature similarity and accuracy of machine learning prediction in GELI condition. There are some basic requirements to apply FSM on CFD modeling and simulation: (1) coarse-mesh CFD simulation can generate reasonable results which capture the basic behaviors of targeted phenomena within an acceptable uncertainty; (2) simulation error is mainly impacted by model error and mesh-induced numerical error where mesh size is one of key model parameters that makes two main error sources tightly connected. (3) training data is qualified (i.e., performance of physical feature) and sufficiently efficient (i.e., size of relevant data for training) for machine learning algorithm to learn from and find the underlying patterns of identified local physical features.

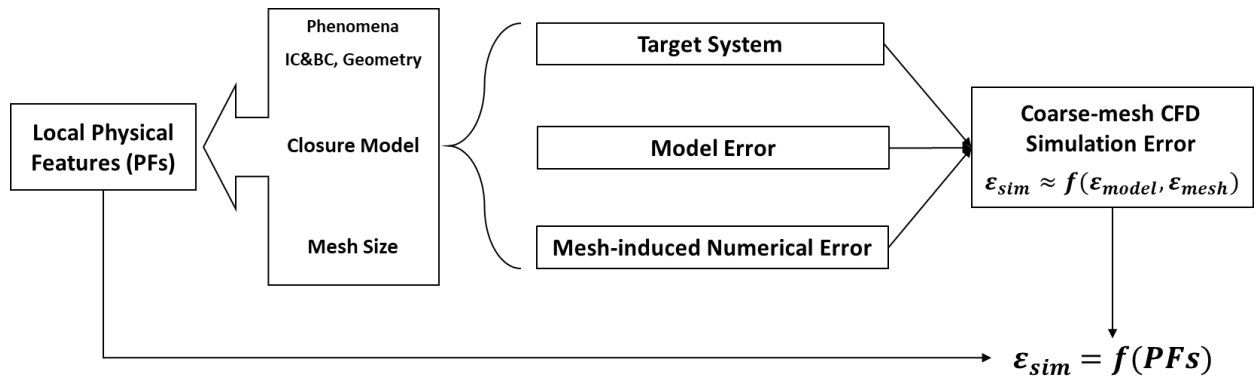

Figure 3. Basic idea of FSM: surrogate modeling of local simulation error and physical features.

By treating main error sources together, the basic idea of FSM is to develop a surrogate model to identify the relationship between simulation error and specific local physical features, as shown in Figure 3. These local physical features are identified based on the system information (e.g., IC/BC, geometry, structure), closure models that contain the information of phenomena of interest and relevant to model error, and local mesh sizes that affect the model error and mesh-induced numerical error. Since the values of physical features are not only determined by mesh sizes but also other physical parameters, the gaps between simulations with different mesh sizes are reduced, which makes it possible to use this well-trained surrogate model to predict the extrapolation of local mesh sizes and use fine-mesh simulation to inform coarse-mesh simulation. Figure 4 displays the workflow of applying FSM for computationally efficient CFD prediction, which is modularized in four independent steps as target analysis, feature identification, training database construction and error prediction. State-of-the-art techniques and algorithms are applied to realize the goals of each step.

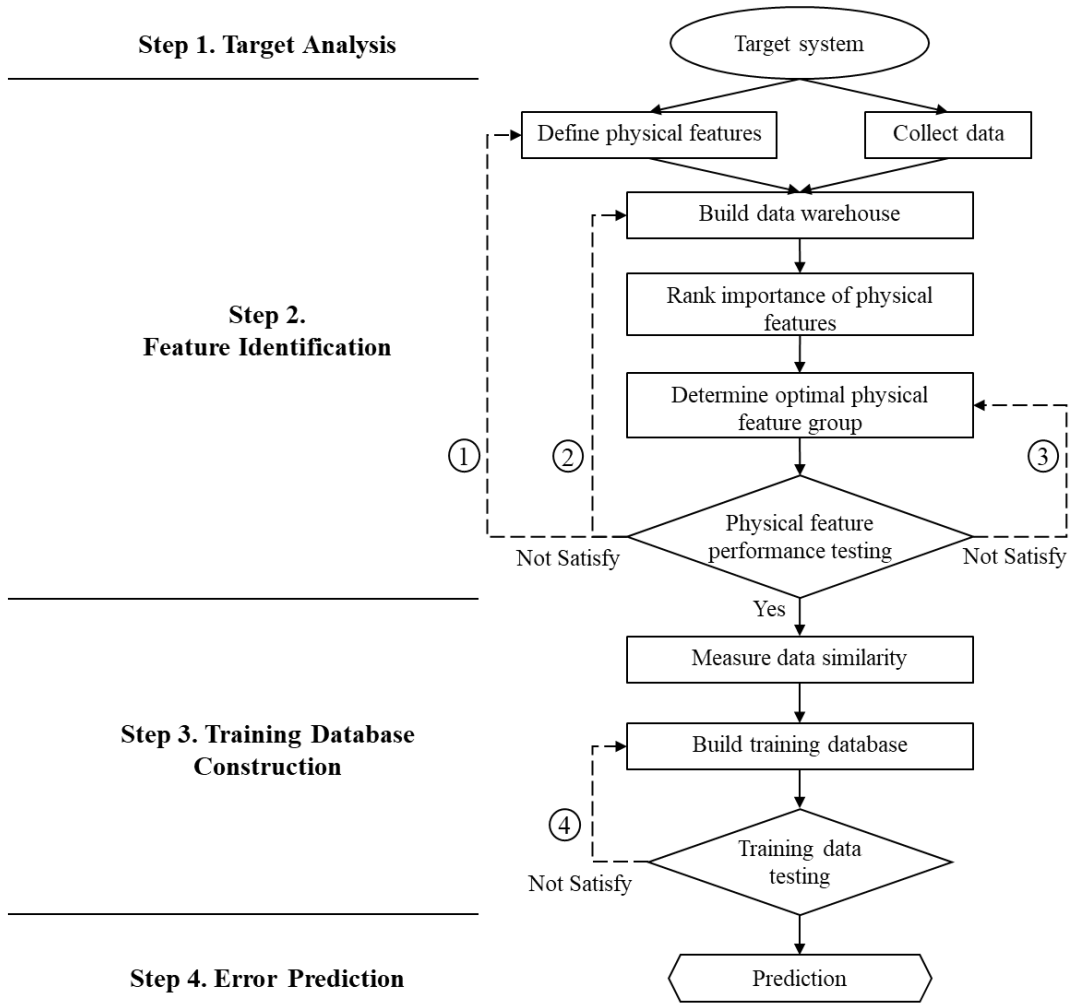

Figure 4. Workflow of applying FSM for computationally efficient CFD prediction.

*3.1. Step 1: target analysis*

This step is to establish the knowledge basis for the ensuing feature identification, training database construction and error prediction. The first thing is to specify the key phenomena involved in the target case. In the system-level thermal-hydraulic simulation of an NPP, the Quantities of Interest (QoIs) are normally influenced by a couple of different phenomena. A PIRT (Phenomenon Identification and Ranking Table) procedure should be executed to decompose the complex physics and identify the key phenomena. For example, a two-phase bubbly flow simulation may imply an interaction of different physical models respectively for two-phase interfacial forces and turbulence. The predictions on the QoIs (e.g., void fraction and velocities) are determined by the closure physical models used for these key phenomena in the simulation tool. Optimizing the prediction accuracy of these key QoIs using coarse-mesh CFD simulation is the purpose of the framework. In addition, some global parameters should be identified to represent the global physical condition of the target case, which helps select training database and construct test matrix. For instance, the Reynolds (Re) number is identified as the key global physical parameter in a fully developed turbulent pipe flow. Finally, a set of coarse mesh sizes should be selected for different control volumes. Based on the capability of simulation tool, these mesh sizes should be in an appropriate range in which they are neither too fine bringing too much computation cost, nor too coarse losing too much local patterns. Overall, the items that should be specified in this step are (1) key phenomena, respective QoIs and applicable closure models for the target case; (2) system information, e.g., geometry, structure and IC/BC; (3) global parameters that represent the global physical condition of the target case; (4) reasonable coarse mesh sizes for CFD simulation.

*3.2. Step 2: feature identification*

This step aims to identify optimal physical feature group for each QoI in three parts: (1) define and quantify potential physical features, (2) analyze and rank physical feature importance for each QoI, (3) test and determine optimal physical feature group for each QoI. Firstly, potential physical features should be defined based on the phenomena, ICs/BCs and geometry of target system, relevant closure models, and local mesh sizes ensuring local physics to be well represented. Then, performing importance analysis to rank the importance of these physical features for each QoI and reducing the dimensionality of physical features. Figure 5 displays the flowchart of Step 1, where $\boldsymbol{PF_N}$ is the potential physical features, $N$ is the number of these physical features, $N_{QoI}$ is the number of QoI. $\boldsymbol{QoI_{LF}}$ and $\boldsymbol{QoI_{HF}}$ are the QoIs calculated from low-fidelity simulation and high-fidelity simulation, $\boldsymbol{\varepsilon} = \boldsymbol{QoI_{HF}} - \boldsymbol{QoI_{LF}}$ is the error between these two simulation results, $[\boldsymbol{S_N}]_1, \ldots, [\boldsymbol{S_N}]_i, \ldots, [\boldsymbol{S_N}]_{N_{QoI}}$ ($1 \leq i \leq N_{QoI}$) are the importance scores obtained by non-parametric importance analysis which represents the importance levels of the physical features for each QoI, $\boldsymbol{PF_{N_1}}, \ldots, \boldsymbol{PF_{N_i}}, \ldots, \boldsymbol{PF_{N_{QoI}}}$ ($\boldsymbol{PF_{N_i}} \in \boldsymbol{PF_N}, 1 \leq i \leq N_{QoI}$) are the optimal physical feature groups for each QoI, the subscripts $\boldsymbol{tr}$ and $\boldsymbol{te}$ means training and testing cases, $\boldsymbol{\varepsilon_{i,all}} = f_i(\boldsymbol{PF_N})$ and $\boldsymbol{\varepsilon_{i,opt}} = g_i(\boldsymbol{PF_{N_i}})$ are the DFNN error models trained by training cases for $QoI_i$ respectively using all defined physical features ($PF_N$) and only optimal physical features ($PF_{N_i}$), $\varepsilon_{i,all,te}$ and $\varepsilon_{i,opt,te}$ are the DFNN error predictions for the $QoI_j$ in the testing cases respectively based on the values of all defined physical features ($PF_{N,te}$) and only optimal physical features ($PF_{N_i,te}$), and $\varepsilon_{i,te}$ is real error of $QoI_i$ between low-fidelity results and high-fidelity results in the testing cases.

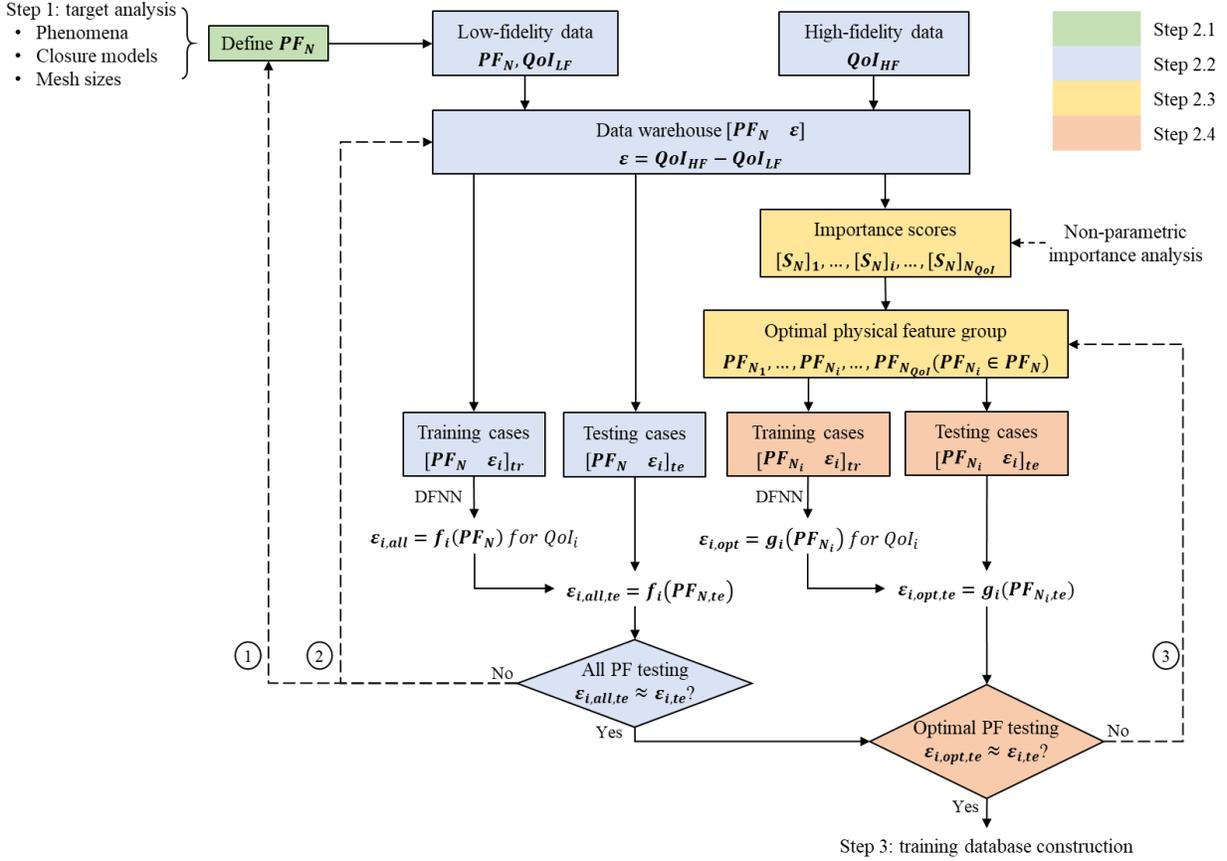

Figure 5. Flowchart of Step 2: feature identification.

### 3.2.1. Step 2.1: define potential physical features

Based on the information about phenomena, closure models and mesh sizes determined in Step 1, potential physical features are defined in a physics-guided way to represent the underlying local patterns of the target system. The local physical features are classified into two types: derivatives of variables that indicate regional information and local physical parameters as shown in Figure 6.

To fully capture the characteristics of local physics, the initial selection of the physical feature should include all the potential ones that satisfy the classification and are relevant to the mesh sizes, phenomena and closure models.

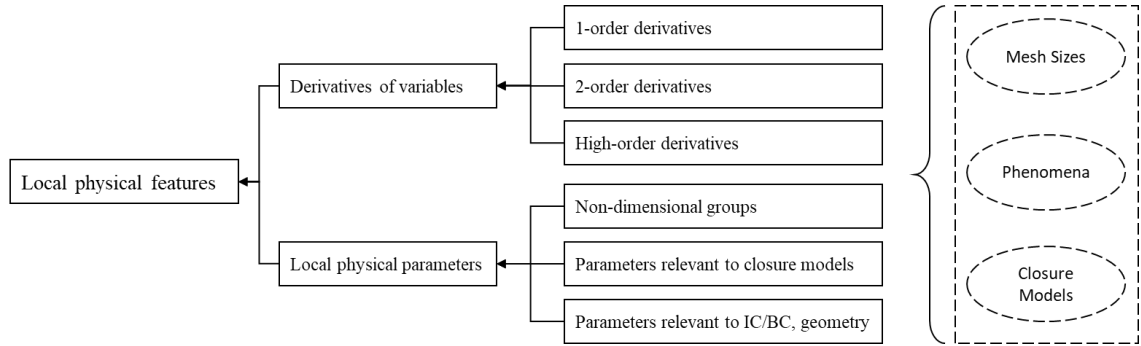

Figure 6. Classification of local physical features.

The derivatives of variables include 1-order, 2-order and high-order derivatives of variables calculated by central-difference formulas, which not only contain the information in local cell but also the information that represents regional physical patterns and connection with adjacent cells. It is analogous to the identification of a person, not only his/her personal information such as height and weight is important, but also his/her connections with other people should be considered. More detailed regional information may be involved if higher-order derivatives are added into the local PF group. The expressions of 1-order and 2-order derivatives of variables in 2D problems are listed in Table 1.

Table 1. Expression of 1-order and 2-order derivatives of variables in 2D problems.

| For cells in bulk | | |
|---|---|---|
| 1-order | $\left.\dfrac{\partial V}{\partial x_i}\right|_{(i,j)} = \dfrac{V_{i+1,j} - V_{i-1,j}}{2\Delta x_i}$ | (10) |
| 2-order | $\left.\dfrac{\partial^2 V}{\partial x_i^2}\right|_{(i,j)} = \dfrac{\left.\dfrac{\partial V}{\partial x_i}\right|_{(i+1,j)} - \left.\dfrac{\partial V}{\partial x_i}\right|_{(i-1,j)}}{2\Delta x_i} = \dfrac{V_{i+2,j} - 2V_{i,j} + V_{i-2,j}}{4(\Delta x_i)^2}$ | (11) |
| | $\left.\dfrac{\partial^2 V}{\partial x_j \partial x_i}\right|_{(i,j)} = \left.\dfrac{\partial^2 V}{\partial x_i \partial x_j}\right|_{(i,j)} = \dfrac{V_{i+2,j+2} - V_{i-1,j+1} - V_{i+1,j-1} + V_{i-2,j-2}}{4\Delta x_i \Delta x_j}$ | (12) |
| For cells adjacent to the wall | | |
| 1-order | $\left.\dfrac{\partial V}{\partial x_i}\right|_{(1,j)} = \dfrac{V_{2,j} - V_{0,j}}{\frac{3}{2}\Delta x_i}$ | (13) |
| 2-order | $\left.\dfrac{\partial^2 V}{\partial x_i^2}\right|_{(1,j)} = \dfrac{\left.\dfrac{\partial V}{\partial x_i}\right|_{(2,j)} - \left.\dfrac{\partial V}{\partial x_i}\right|_{(0,j)}}{\frac{3}{2}\Delta x_i} = \dfrac{V_{3,j} - 5V_{1,j} + 4V_{0,j}}{3(\Delta x_i)^2}$ | (14) |
| | $\left.\dfrac{\partial^2 V}{\partial x_i \partial x_j}\right|_{(1,j)} = \left.\dfrac{\partial^2 V}{\partial x_j \partial x_i}\right|_{(1,j)} = \dfrac{V_{2,j+1} - V_{2,j-1} - V_{0,j+1} + V_{0,j-1}}{3\Delta x_i \Delta x_j}$ | (15) |

3.2.2. Step 2.2: collect data and build data warehouse

All available high-fidelity data that is relevant to the phenomena involved in the target case should be collected and processed to build the data warehouse, which includes experimental observation, DNS data, and validated high-resolution numerical results. The definition of "high-fidelity" depends on the requirements of simulation accuracy for the target case. According to the physical conditions of limited high-fidelity data, low-fidelity data can be collected or generated using CFD codes with coarse mesh sizes and closure models identified in Step 1. As shown in Figure 5, $QoI_{LF}$ and $PF_N$ are calculated based on low-fidelity simulation results while $QoI_{HF}$ is from high-fidelity data. Then the data warehouse is built including physical features and simulation errors of local variables as $[PF_N \quad \varepsilon]$, where $\varepsilon = QoI_{HF} - QoI_{LF}$. There are two methods to calculate the error between fine-mesh high-fidelity data and coarse-mesh low-fidelity data: point-to-point and cell-to-cell. The point-to-point method compares the values of local variables at the exact locations existing in both of high-fidelity and low-fidelity data. This method can be applied if both high-fidelity and low-fidelity simulations are using the finite element method or the finite difference method. The cell-to-cell method compares the values of local variables in the coarse-mesh cell by averaging and mapping the high-fidelity data from fine cells to coarse ones.

Then the performance of $PF_N$ should be tested to see whether they can represent local physics and provide sufficiently accurate prediction on simulation error of $QoI_i$. By randomly dividing the data sets into two parts as training cases and testing cases, the DFNN surrogate model can be trained and fitted as $\varepsilon_{i,all} = f_i(PF_N)$ for $QoI_i$ using $PF_{N,tr}$ as inputs and $\varepsilon_{i,tr}$ as outputs. Inserting the values of $PF_{N,te}$ into this model, DFNN prediction can be obtained as $\varepsilon_{i,all,te} =$ $f_i(\boldsymbol{PF}_{N,te})$. The performance of $\boldsymbol{PF}_N$ can be quantified as the Normalized Root Mean Squared Error (NRMSE) between $\boldsymbol{\varepsilon}_{i,all,te}$ and the $\boldsymbol{\varepsilon}_{i,te}$,

$$NRMSE_{i,all} = \frac{\sqrt{\frac{1}{n}\sum(\boldsymbol{\varepsilon}_{i,all,te} - \boldsymbol{\varepsilon}_{i,te})^2}}{\frac{1}{n}\sum \boldsymbol{\varepsilon}_{i,te}} \quad (16)$$

where n is the number of testing data points. If the value of $NRMSE_{i,all}$ is sufficiently accurate and satisfy the simulation requirement, these physical features have the capability to capture the local physics. Otherwise, two ways of improvement are available which are denoted as dashed lines in both Figure 4 and Figure 5: (1) defining more physical features and (2) collecting more relevant data.

### 3.2.3. Step 2.3: rank importance of physical features

Depending on the complexity of phenomena, dimensionality of simulation and system conditions, the number of potential physical features defined in Step 2.1 may be large, which increases the computational cost in data processing and training of DFNN models. The step 2.3 aims to apply non-parametric importance analysis for the dimensionality reduction of physical feature, by which to choose the optimal physical feature groups for different QoIs and keep a balance between computational costs and prediction accuracy. In this paper, a statistical technique, random forest regression (RF regression, or RFR) (Breiman, 2001) is applied to quantify and rank the importance of physical features. RFR is applicable for pure data, since the data of input variables can be generated by calling the response function or sampling from a prepared database. As a non-parametric method, RFR does not require a fixed regression model form or an uncorrelated relationship between the input variables, which makes it suitable here considering the highly non-linear relationship between physical features and local simulation errors. Same as DFNN, RFR is also a supervised learning algorithm but much more computationally efficient, therefore, it can also be used as a fast-running tool to pre-evaluate the predictive performance of these defined physical features.

As an ensemble learning technique, RFR works by constructing a forest of uncorrelated regression trees at training time and outputting a mean prediction from these individual trees. Bootstrap aggregating (or bagging) technique is applied as training algorithm for random forests.

Once the regression trees have been built, the importance of variables can be measured by observing the Out-Of-Bag (OOB) error, which is called the Permutation Variable Importance Measure (PVIM) (Breiman, 2001). The following process describes the estimation of variable importance values by PVIM. Suppose the OOB data can be expressed as $B_m = \{(y_j^m, x_j^m), m = 1,2,...M \text{ and } j = 1,2,...,S\}$, where $S$ is the number of sample points.

1. For the $m$th tree, the prediction errors on the OOB data before and after randomly permuting the values of the input variable $X_f$ $(f = 1,2,...,F)$ are calculated using,

$$MSE_m = \frac{1}{S}\sum_{j=1}^{S}(y_j^m - \hat{y}_j^m)^2 \text{ and } MSE_{m,f} = \frac{1}{S}\sum_{j=1}^{S}(y_j^m - \hat{y}_{j,f}^m)^2 \quad (17)$$

where $\hat{y}_j^m$ and $\hat{y}_{j,f}^m$ are the prediction from the $m$th tree respectively before and after permutation.

2. The differences between two predictions are defined as the value of PVIM:

$$PVIM_{m,f} = MSE_{m,f} - MSE_m \quad (18)$$

3. The overall PVIM of $X_f$ in the OOB data is then calculated as

$$PVIM_f = \frac{\frac{1}{M}\sum_{m=1}^{M} PVIM_{m,f}}{\sigma_f} \quad (19)$$

where $\sigma_f$ is the standard deviation of the differences over the total OOB data. The value of $PVIM_f$ indicates the OOB importance of $X_f$ on the response. In this way, the OOB importance can be measured for each input variable. In the $m$th tree, if $X_f$ is not selected as the splitting variable, then $PVIM_f = 0$. This implies that the interactions between $X_f$ and other variables are considered to measure its contribution on the prediction accuracy. The importance of a variable increases with the value of PVIM. Therefore, the importance of each physical feature for each QoI can be quantified with the values of PVIM as $[S_N]_1, ..., [S_N]_i, ..., [S_N]_{N_{QoI}}$ and can be ranked in several levels. Based on the importance levels, different physical feature groups for each QoI can be generated respectively including physical features in different importance levels, the predictive capability of which will be evaluated in the Step 2.4 to determine which group is the optimal one for the specific QoI.

3.2.4. Step 2.4: determine optimal physical feature group

After the number of physical features for $QoI_i$ is initially reduced from $\boldsymbol{PF_N}$ to $\boldsymbol{PF_{N_i}}$, the computational cost is reduced in the DFNN training but uncertainty is introduced due to dimensionality reduction of physical feature. The initially selected optimal physical feature group $\boldsymbol{PF_{N_i}}$ may be not sufficient to represent the underlying local physics as $\boldsymbol{PF_N}$. Therefore, it is necessary to re-test the predictive capability of $\boldsymbol{PF_{N_i}}$. The selection of an optimal physical feature group for $QoI_i$ should provide the required accuracy with minimal computational cost, where these requirements are defined according to the application. Two metrics should be considered here to determine the optimal $\boldsymbol{PF_{N_i}}$: (1) NRMSE of prediction to evaluate whether the reduced physical feature group keeps the underlying physics, and (2) Computational cost for data training to find how much computation is saved after the dimensionality reduction. The same training cases and testing cases in Step 2.2 are used for this testing to compare the training times and NRMSEs between $\boldsymbol{PF_N}$ and $\boldsymbol{PF_{N_i}}$,

$$NRMSE_{i,opt} = \frac{\sqrt{\frac{1}{n}\sum(\varepsilon_{i,opt,te} - \varepsilon_{i,te})^2}}{\frac{1}{n}\sum \varepsilon_{i,te}} \qquad (20)$$

*3.3. Step 3: training database construction*

This step aims to answer the question: which kind of data in the data warehouse should be used as the training database? Training database should be sufficient to capture the local physics in the target case, and efficient to avoid huge computational cost on data training and processing. Therefore, another question needs to be answered: how to determine the similarity of target data and training data? One of the claims is that if target data is more like training data, machine learning prediction error on the target case is smaller. Considering these target-similar data points distribute in different cases in training data warehouse, the optimal training database should be constructed by choosing these target-similar data points in one case instead of choosing the entire case. Figure 7 illustrates the information flow of Step 3.

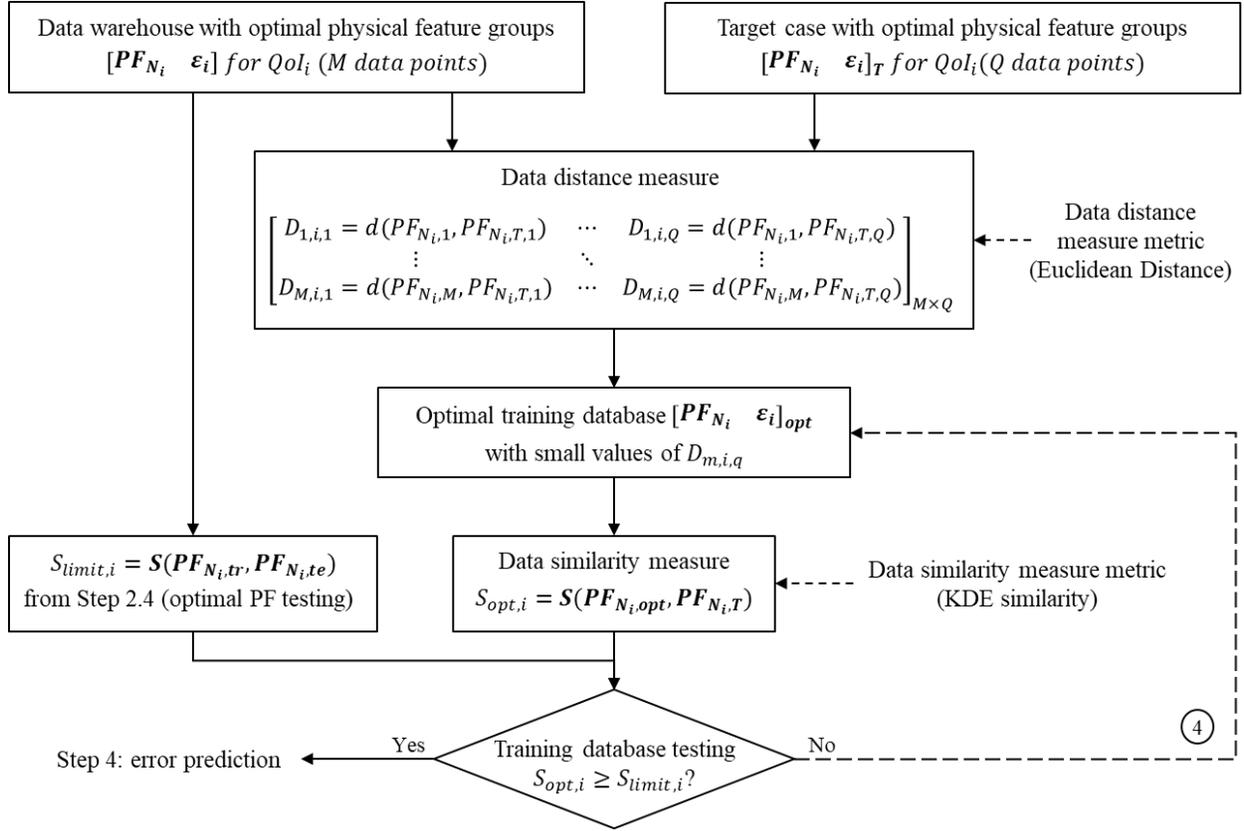

Figure 7. Flowchart of Step 3: training database construction.

Provided that the number ($Q$) of target data points is much smaller than the number ($M$) of data points in the data warehouse, the target-similar data points can be determined by measuring their data distance with each target data points. Euclidean distance is used as a metric to calculate the distance between single data points, as expressed as below,

$$D_{m,i,q} = d(PF_{N_i,m}, PF_{N_i,T,q}) = \sqrt{\sum_{k=1}^{N_i}(x_{tr,k} - x_{ta,k})^2} \qquad (21)$$

$$(1 \leq m \leq M \text{ and } 1 \leq q \leq Q)$$

For each target data point $[PF_{N_i} \quad \varepsilon_i]_{T,q}$, the data points with small values of Euclidean distance can be identified. Therefore, the optimal training database $[PF_{N_i} \quad \varepsilon_i]_{opt}$ can be constructed by including these target-similar data points. The data similarity between $[PF_{N_i} \quad \varepsilon_i]_{opt}$ and target case $[PF_{N_i} \quad \varepsilon_i]_T$ can be measured using data similarity measure metrics. In this paper, a method called Kernel Density Estimation (KDE) is introduced to develop a measure metric. As a non-parametric way to estimate the probability density function, KDE assumes the data distribution can be approximated as a sum of multivariate Gaussians. A kernel distribution can be used if a parametric distribution cannot properly describe the data, or to avoid making assumptions about the distribution of the data. KDE can be considered as the probability that the data point ($q$) locates in the distribution of training data ($p_i, i = 1,2, \dots, N$) as (Scott, 2015),

$$p_{KDE} = \frac{1}{N \cdot h_1 h_2 \dots h_d} \sum_{i=1}^{N} \prod_{j=1}^{d} ker(\frac{q_j - p_{i,j}}{h_j}) \qquad (22)$$

where $d$ is the number of variables in $q$ and $p_i$. $ker$ is the kernel smoothing function. $h_j$ is the bandwidth for each variable. A multivariate kernel distribution is defined by a smoothing function ($k$) and a bandwidth matrix defined by $H = h_1, h_2, \dots, h_d$, which control the smoothness of the resulting density curve. Therefore, KDE method can be used to measure the data similarity by estimating the probability of a given point locating in a set of training data points. In this step, the similarity between training data ($p_i, i = 1,2, \dots, N$) and target data ($q_j, k = 1,2, \dots, M$) is expressed as the mean of KDEs below,

$$S_{KDE} = \frac{1}{M} \sum_{k=1}^{M} p_{KDE,k} = \sum_{k=1}^{M} \frac{1}{N \cdot h_1 h_2 \dots h_d} \sum_{i=1}^{N} \prod_{j=1}^{d} ker(\frac{q_{j,k} - p_{i,j}}{h_j}) \qquad (23)$$

A greater value of $S_{KDE}$ means higher level of similarity. To ensure the optimal training database has sufficient predictive capability for $QoI_i$, a test should be performed to evaluate whether this level of data similarity is high enough compared with the limit value $S_{limit,i}$. For different phenomena, geometry, and simulation tools, the value of $S_{limit,i}$ is different so it is difficult to provide a fixed general value. In Step 2.4 of this framework, a test (with training data $[PF_{N_i} \quad \varepsilon_i]_{tr}$ and testing data $[PF_{N_i} \quad \varepsilon_i]_{te}$) has been developed for optimal physical feature testing. The value of data similarity between $PF_{N_i,tr}$ and $PF_{N_i,te}$ is set as $S_{limit,i}$ since all the relatively irrelevant data points in data warehouse are also included in the training database, which lowers the level of similarity. If the level of similarity does not satisfy the requirement, a feedback is generated for the re-selection of optimal training database, denoted as dashed line 4 in both Figure 4 and Figure 7.

*3.4. Step 4: error prediction for coarse-mesh CFD*

After determining optimal physical features and training database, the error prediction for coarse-mesh CFD simulation can be implemented for the target case.

## 4. Demonstration of the framework for two-phase bubbly flow

In this paper, a case study based on two-phase bubbly flow was performed to evaluate the predictive capability of FSM in two-phase flow coarse-mesh CFD simulation which adopts the Eulerian-Eulerian two-fluid model as discussed in the Appendix. It is upward bubbly pipe flow experiments for water at atmospheric pressure and temperature of 10 $^{0}$C without phase change. 42 reference experimental datasets with different injection rates and void fractions are from Liu and Bankoff (Liu and Bankoff, 1993) and used for the validation of high-fidelity simulation data. **Error! Reference source not found.** lists the global injection conditions of the 42 experimental cases. Both high-fidelity and low-fidelity simulations were performed using the commercial CFD package, STAR-CCM+12.06, with the following two-phase interfacial forces closures and turbulence models: drag force model from Tomiyama (Tomiyama et al., 1998), lift correction from Shaver and Podowski (Shaver and Podowski, 2015) with a base coefficient of 0.025, turbulent dispersion force from Burns (Burns et al., 2004) and the standard k-ε turbulence model (Jones and Launder, 1972). The set of these models which is referred as the Bubbly and Moderate Void Fraction (BAMF) model has been tested and validated for 12 cases from the Liu and Bankoff experimental datasets, which provided reasonable predictions for mean flow profiles of void fraction and phase velocities (Sugrue et al., 2017). In this paper, all the experimental datasets are simulated with the BAMF model to gain sufficient database.

Table 2. Summary of the flow conditions of experimental cases.

| Set | Liquid rate (m/s) | Vapor rate (m/s) | Void fraction | Set | Liquid rate (m/s) | Vapor rate (m/s) | Void fraction |
|---|---|---|---|---|---|---|---|
| 1 | 0.376 | 0.027 | 0.0407 | 22 | 0.974 | 0.027 | 0.0204 |
| 2 | 0.376 | 0.067 | 0.1167 | 23 | 0.974 | 0.067 | 0.0514 |
| 3 | 0.376 | 0.112 | 0.1843 | 24 | 0.974 | 0.112 | 0.0791 |
| 4 | 0.376 | 0.18 | 0.2449 | 25 | 0.974 | 0.18 | 0.1242 |
| 5 | 0.376 | 0.23 | 0.3079 | 26 | 0.974 | 0.23 | 0.1512 |
| 6 | 0.376 | 0.293 | 0.3657 | 27 | 0.974 | 0.293 | 0.1869 |
| 7 | 0.376 | 0.347 | 0.4168 | 28 | 0.974 | 0.347 | 0.2108 |
| 8 | 0.535 | 0.027 | 0.0312 | 29 | 1.087 | 0.027 | 0.0176 |
| 9 | 0.535 | 0.067 | 0.0877 | 30 | 1.087 | 0.067 | 0.0473 |
| 10 | 0.535 | 0.112 | 0.1406 | 31 | 1.087 | 0.112 | 0.0737 |
| 11 | 0.535 | 0.18 | 0.2016 | 32 | 1.087 | 0.18 | 0.1096 |
| 12 | 0.535 | 0.23 | 0.2344 | 33 | 1.087 | 0.23 | 0.1497 |
| 13 | 0.535 | 0.293 | 0.3102 | 34 | 1.087 | 0.293 | 0.1777 |
| 14 | 0.535 | 0.347 | 0.3398 | 35 | 1.087 | 0.347 | 0.1976 |
| 15 | 0.753 | 0.027 | 0.0235 | 36 | 1.391 | 0.027 | 0.0148 |
| 16 | 0.753 | 0.067 | 0.0622 | 37 | 1.391 | 0.067 | 0.0387 |
| 17 | 0.753 | 0.112 | 0.1091 | 38 | 1.391 | 0.112 | 0.0581 |
| 18 | 0.753 | 0.18 | 0.1554 | 39 | 1.391 | 0.18 | 0.0964 |
| 19 | 0.753 | 0.23 | 0.1816 | 40 | 1.391 | 0.23 | 0.1176 |
| 20 | 0.753 | 0.293 | 0.2381 | 41 | 1.391 | 0.293 | 0.1504 |
| 21 | 0.753 | 0.347 | 0.2692 | 42 | 1.391 | 0.347 | 0.1724 |

To accelerate the simulation, as shown in Figure 8, only one-quarter of the domain is simulated, and symmetric boundary conditions are applied on the two side surfaces. For each of the 42 experimental cases, four sets of mesh configurations are selected and the cross sectional view of the mesh configurations are shown in Figure 8, where the mesh configuration of the reference high-fidelity case is identical to the cases reported in (Sugrue et al., 2017) and the other three low-fidelity cases adopt coarse mesh resolutions to analyze the integration of mesh-induced numerical and physical model error. The numbers of cells from the wall to the pipe center are 10, 15, 20 and 25, thus having the total number of cells in the domain as 0.07, 0.18, 0.35, and 0.63 million.

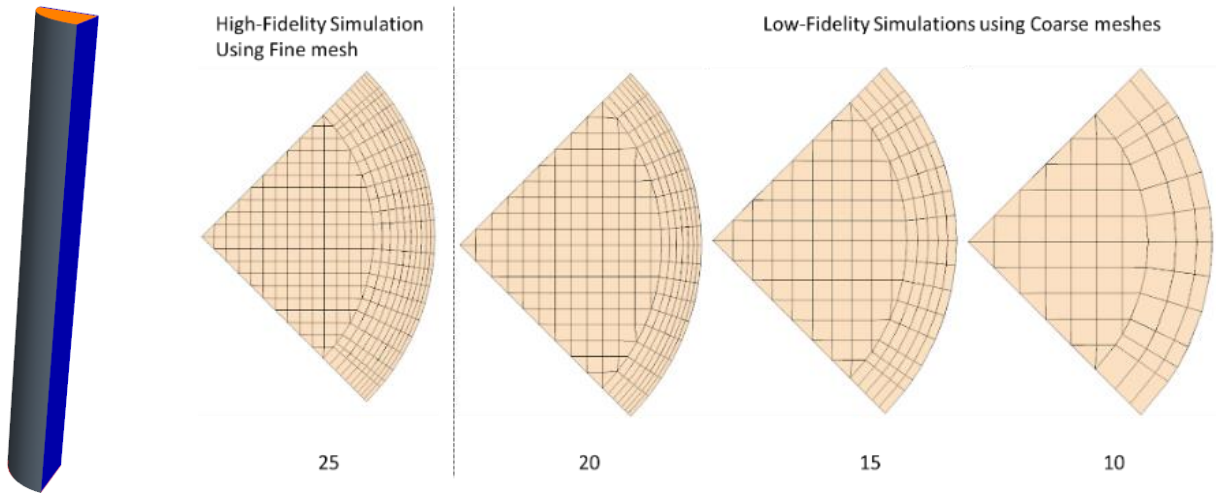

Figure 8. Cross-sectional view of the mesh configuration for high and low fidelity simulations.

The application of FSM on bubbly flow coarse-mesh CFD simulation is described in this section. Of all the 42 cases, Case 35 can be considered as an extrapolation of vapor injection rate. It also has the highest simulation errors between high-fidelity simulation and low-fidelity simulations. Therefore, the goal of this case study is to predict the low-fidelity simulation errors of Case 35 using part of other 41 cases for data training.

*4.1. Step 1: target analysis*

As described in problem statement, the key phenomena involved in the target case is bubbly pipe flow. The following closure models are applied: (1) two-phase interfacial forces closures including drag force model and lift correction, and (2) turbulence models including turbulent dispersion force and the standard k-ε turbulence model. QoIs in this simulation are liquid velocity, vapor velocity and void fraction. Global parameters that represent the global physical condition of the target case are liquid and vapor injection rates and injection void fraction. Three different coarse mesh sizes are applied for low-fidelity simulations, the total number of cells is reduced respectively to 11%, 29% and 56% of that of fine-mesh simulation. The computational cost using 10 cells is significantly saved compared with others, thus in this case study, the error prediction is performed for the simulation using 10 cells.

*4.2. Step 2: feature identification*

### 4.2.1. Step 2.1: define potential physical features

According to the involved phenomena, applied closure models, IC/BC and geometry information, 27 potential local physical features are defined. 16 of all are 1-order and 2-order derivatives of variables, which include liquid and vapor velocity ($u_l$ and $u_g$), void fraction ($\alpha$), pressure ($P$), liquid and vapor kinetic energy ($k_l$ and $k_g$), liquid and vapor turbulence dissipation rate ($\varepsilon_l$ and $\varepsilon_g$). A set of non-dimensional parameters which are adopted and customized to characterize the flow features is included as well. The principle is to include as many relevant non-dimensional parameters as possible while the importance of each parameter will be justified by the machine learning algorithms. Defined as the ratio between the inertial forces to viscous forces, in this work, Reynolds number is customized to have three different expressions representing local flow features. Three different local Reynolds number $Re_\Delta$, $Re_b$, and $Re_y$ respectively use local mesh size ($\Delta$), pre-set bubble size ($D_b$) and wall distance ($y$) as characteristic lengths, which take the effects of mesh, closure model and geometry into consideration. Weber number characterizes the relative importance of the fluid's inertia compared to its surface tension. Turbulent intensity ($I_l$ and $I_g$) provides a measurement of the flow fluctuations versus the mean flow velocity. $R_l$ and $R_g$ are defined as the ratio between turbulent length scale, i.e., $\frac{k^{\frac{3}{2}}}{\varepsilon}$, and the bubble diameter, $D_b$. Non-dimensional wall distance, $R_b$, is included as well since the velocity and void fraction distributions crucially depend on the wall distance. $r_l$ represents the ratio between the turbulent eddy viscosity of liquid and the molecular viscosity of liquid which is supposed to become important for the high Reynolds number regime. $R_\mu$ represents the ratio between the gas and liquid eddy viscosity which characterizes the magnitude of modeled turbulence level for liquid and gas.

Table 3. Identification of physical feature ($PF_N$) for bubbly flow CFD simulation.

| Derivatives of variable | | | | Local physical parameters | | | | | |
|---|---|---|---|---|---|---|---|---|---|
| 1-order derivatives | | 2-order derivatives | | Non-dimensional groups | | | Parameters relevant to closure models, IC/BC, geometry | | |
| $\dfrac{du_l}{dx}$ | $\dfrac{du_g}{dx}$ | $\dfrac{d^2 u_l}{dx^2}$ | $\dfrac{d^2 u_g}{dx^2}$ | $Re_\Delta = \dfrac{\rho_l \Delta \cdot \Delta u}{\mu_l}$ | $I_l = \dfrac{k_l}{u_l^2}$ | | $R_l = \dfrac{k_l^{\frac{3}{2}}}{\varepsilon_l D_b}$ | | $R_\mu = \dfrac{\mu_g^t}{\mu_L^t}$ |
| $\dfrac{d\alpha}{dx}$ | $\dfrac{dP}{dx}$ | $\dfrac{d^2\alpha}{dx^2}$ | $\dfrac{d^2 P}{dx^2}$ | $Re_b = \dfrac{\rho_l D_b \Delta u}{\mu_l}$ | $I_g = \dfrac{k_g}{u_g^2}$ | | $R_g = \dfrac{k_g^{\frac{3}{2}}}{\varepsilon_g D_b}$ | | $r_l = \dfrac{\mu_l^t}{\mu_l}$ |
| $\dfrac{dk_l}{dx}$ | $\dfrac{dk_g}{dx}$ | $\dfrac{d^2 k_l}{dx^2}$ | $\dfrac{d^2 k_g}{dx^2}$ | $We = \dfrac{\rho D_b \Delta u^2}{\sigma}$ | | | $Re_y = \dfrac{\rho_l y \Delta u}{\mu_l}$ | | $R_b = \dfrac{D_b}{\Delta}$ |
| $\dfrac{d\varepsilon_l}{dx}$ | $\dfrac{d\varepsilon_g}{dx}$ | $\dfrac{d^2 \varepsilon_l}{dx^2}$ | $\dfrac{d^2 \varepsilon_g}{dx^2}$ | | | | | | |

4.2.2. Step 2.2: collect data and build data warehouse

Using the BAMF model, high-fidelity and low-fidelity data were generated by STAR-CCM+12.06 with fine mesh and three different coarse meshes respectively. The point-to-point method is applied to calculate the simulation errors of local QoIs ($u_l, u_g, \alpha$) in this case study. Then the performance of $PF_N$ is tested using Case 34 as the testing case and other 40 cases as training cases. With 27 inputs and 3 outputs, a DFNN containing 3 hidden layers and 30 neurons in each hidden layer (i.e., 20-20-20 DFNN) is applied for data training and simulation error prediction. Figure 9 shows the comparisons between original low-fidelity simulation results and modified results based on the DFNN simulation error prediction for Case 34. Table 4 lists the NRMSEs of DFNN predictions and low-fidelity simulations for Case 34. The DFNN predicted results have smaller NRMSEs than the original low-fidelity simulation results for all the QoIs. After being corrected by the predicted simulation errors from well-trained DFNN model, the accuracy of coarse-mesh simulation results is improved. The original coarse-mesh low-fidelity results using 10-cell configuration show different patterns with the high-fidelity data, there are unphysical "peaks" locating at the ninth point. After the training, the DFNN model well captures the pattern of these unexpected "peaks" and provides an appropriate correction to match the high-fidelity results. The capability of capturing reginal patterns results from identifying 1-order and 2-order derivatives of QoIs as the physical features, because not only the characteristics at this point are captured, but also the connections of this point with its neighboring points. The results indicate that these defined local physical features can represent local physics and provide sufficiently accurate prediction on simulation error of QoIs. FSM represents good predictive capability on estimating the local simulation error even for the extrapolation of global physics (vapor injection rate in this test).

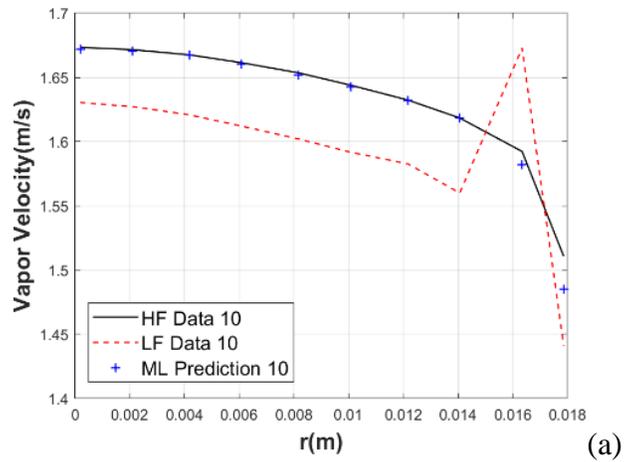

(a)

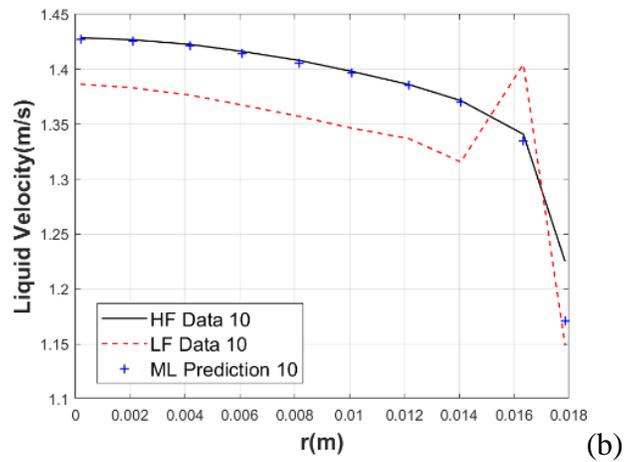

(b)

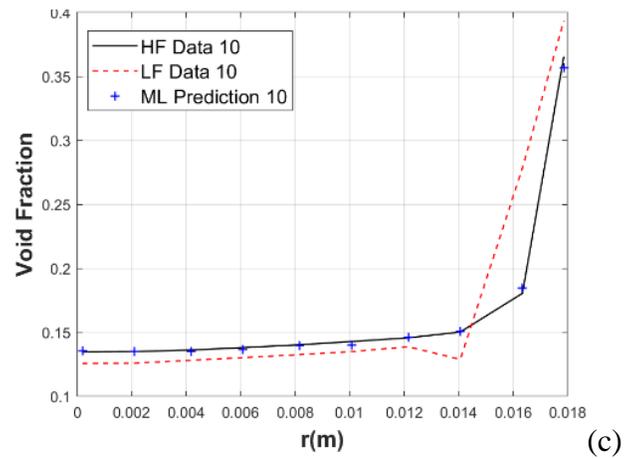

(c)

Figure 9. Comparisons between original low-fidelity simulation results and modified results based on the DFNN simulation error prediction for Case 34.

Table 4. NRMSEs of DFNN predictions and low-fidelity simulations for Case 34.

| Testing case | Testing cases | $NRMSE_{u_g}$ | $NRMSE_{u_l}$ | $NRMSE_\alpha$ | Mesh configuration |
|---|---|---|---|---|---|
| 34 | 1~33 and 36~42 | 0.0054 | 0.0079 | 0.0257 | 10 cells |
| Original low-fidelity simulation | | 0.0341 | 0.0389 | 0.2013 | |

4.2.3. Step 2.3: rank importance of physical features

By applying RFR-PVIM algorithm introduced in Section 3.2.3, the importance scores of all defined potential physical features for each QoI are quantified and ranked, as shown in Figure 10. A greater score implies higher level of importance. According to their importance scores for each QoI, physical features are classified in four levels: level 1 (score ≥ 0.2), level 2 (0.2 > score ≥ 0.15), level 3 (0.15 > score ≥ 0.1), level 4 (0.1 > score ≥ 0).

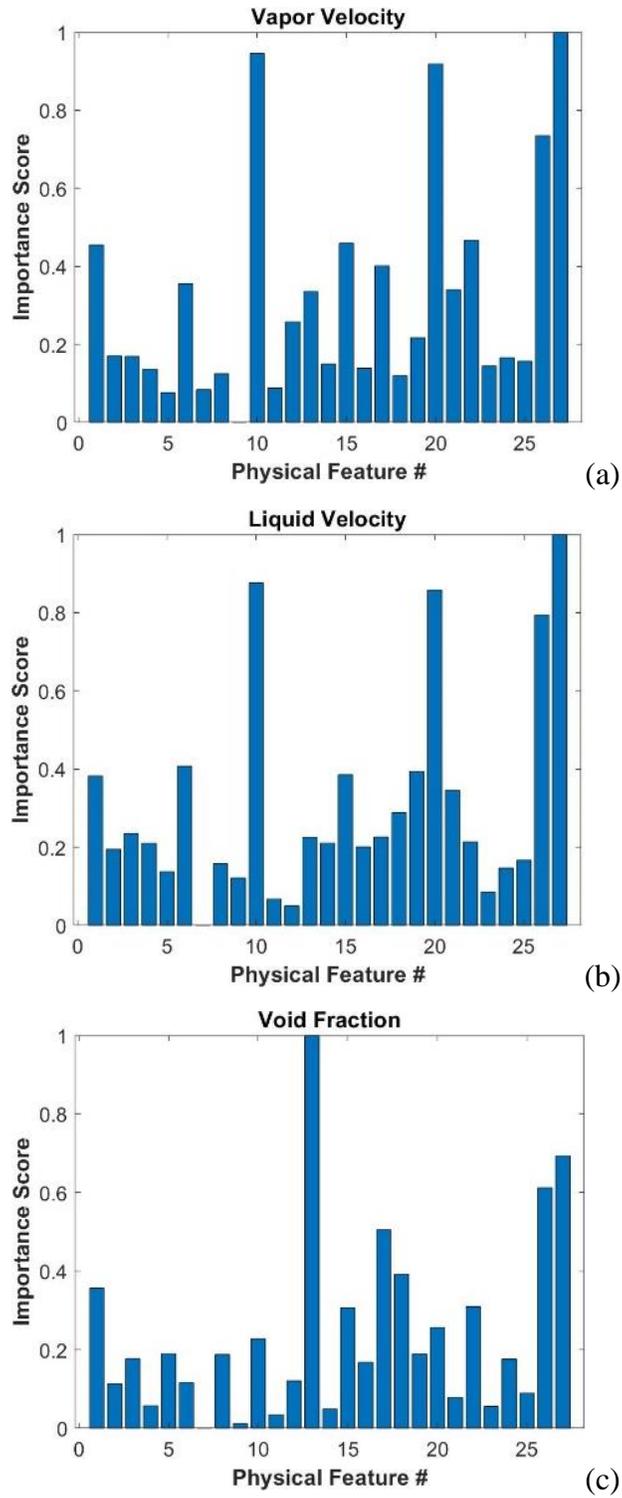

Figure 10. Importance estimation of physical features on different QoIs using RFR-PVIM.

4.2.4. Step 2.4: determine optimal physical feature group

For the DFNN model training and QoI prediction, four different physical feature group can be generated respectively including physical features in level 1, level 1~2, level 1~3 and level 1~4. The prediction errors and computational costs for data training of these four physical feature groups for each QoI are listed in Table 5. For the simulation error estimation of vapor and liquid velocities, physical feature Group 2 with importance level 1 and 2 shows better predictive performance than other groups and has a relatively short training time. For the simulation error estimation of void fraction, physical feature Group 3 with importance level 1 and 2 performs a better balance between prediction accuracy and training cost than other groups. Therefore, these groups are selected as the optimal physical feature groups for these QoIs, as shown in Table 6. More fine levels can be defined to generate more groups if needed, and predictive capability can be improved using complex DFNN structures. All these physical feature groups have the capability to improve the low-fidelity simulations. The final optimal physical feature groups for each QoI are listed in Table 6.

Table 5. Predictive capability and computational cost of different physical feature groups for the test (testing case: Case 34; training cases: 1~33 and 36~42; using a 20-20-20 DFNN).

| QoI | Group # | Number of physical features | $NRMSE$ (10 cells) | Training Time (core-hours) |
|---|---|---|---|---|
| $u_g$ | G1 (Level 1) | 13 | 0.0108 | 0.35 |
| | G2 (Level 1~2) | 17 | 0.0046 | 0.45 |
| | G3 (Level 1~3) | 22 | 0.0076 | 0.85 |
| | G4 (Level 1~4, all) | 27 | 0.0054 | 1.50 |
| | Original low-fidelity simulation | | 0.0341 | |
| $u_l$ | G1 (Level 1) | 17 | 0.0281 | 0.45 |
| | G2 (Level 1~2) | 20 | 0.0050 | 0.50 |
| | G3 (Level 1~3) | 23 | 0.0107 | 0.70 |
| | G4 (Level 1~4, all) | 27 | 0.0079 | 1.50 |
| | Original low-fidelity simulation | | 0.0389 | |
| $\alpha$ | G1 (Level 1) | 10 | 0.1405 | 0.60 |
| | G2 (Level 1~2) | 16 | 0.0744 | 0.85 |
| | G3 (Level 1~3) | 19 | 0.0301 | 0.90 |

| | | | |
|---|---|---|---|
| G4 (Level 1~4, all) | 27 | 0.0257 | 1.50 |
| Original low-fidelity simulation | | 0.2013 | |

Table 6. The optimal physical feature group for the prediction of QoIs.

| QoI | Optimal physical feature | Number |
|---|---|---|
| $u_g$ | $\frac{d\alpha}{dx}, \frac{du_g}{dx}, \frac{dk_g}{dx}, \frac{d^2\alpha}{dx^2}, \frac{d^2P}{dx^2}, \frac{d^2k_g}{dx^2}, \frac{d^2k_l}{dx^2}, \frac{d^2\varepsilon_l}{dx^2}$<br>$I_l, I_g, , We, Re_\Delta, R_b, R_g, R_l, R_\mu, r_l$ | 17 |
| $u_l$ | $\frac{d\alpha}{dx}, \frac{du_g}{dx}, \frac{du_l}{dx}, \frac{dP}{dx}, \frac{d\varepsilon_g}{dx}, \frac{d^2\alpha}{dx^2}, \frac{d^2u_g}{dx^2}, \frac{d^2k_g}{dx^2}, \frac{d^2k_l}{dx^2}, \frac{d^2\varepsilon_l}{dx^2}, \frac{d^2P}{dx^2}$<br>$I_l, We, Re_b, Re_\Delta, R_b, R_g, R_l, R_\mu, r_l$ | 20 |
| $\alpha$ | $\frac{d\alpha}{dx}, \frac{du_g}{dx}, \frac{du_l}{dx}, \frac{dk_g}{dx}, \frac{dk_l}{dx}, \frac{d^2\alpha}{dx^2}, \frac{d^2u_g}{dx^2}, \frac{d^2k_g}{dx^2}, \frac{d^2k_l}{dx^2}, \frac{d^2\varepsilon_l}{dx^2}, \frac{d^2P}{dx^2}$<br>$I_l, I_g, We, Re_b, Re_\Delta, R_b, R_l, R_\mu$ | 19 |

*4.3. Step 3: training data base construction*

Firstly, the data distances between each data points in the data warehouse and the target case need to be calculated and ranked. To ensure the quantity of training database is sufficient, 100 data points with small values of data distance are selected for each target data point. Therefore, there are 1000 data points used to build the training database of a 10-cell coarse-mesh configuration. The weight of data points is considered since some data points are similar to more than one target data points and counted more than one time. The information of the optimal training database for each QoI is summarized in Table 7. It should be explained here that $S_{KDE}$ only denotes the relative data similarity of different database with the target data for one specific QoI. Even $S_{KDE,\alpha}$ has a greater value than $S_{KDE,u_g}$, it doesn't mean the error prediction of $\alpha$ is more accurate than $u_g$.

The results in Table 7 show that the 10-cell and 15-cell data points have higher level of similarity than 20-cell data points since nearly all selected data points come from 10-cell and 15-cell simulations, even there are more data points from 20-cell simulations. For each QoI, more than half of the cases are involved in the training data selection, even the global conditions (e.g., $u_g$, $u_l$ and $\alpha$) of some cases are quite different from those of Case 35.

Table 7. Summary of optimal training database for 10-cell simulation error prediction (Case 35).

| QoI | Data similarity ($S_{KDE}$) | Data source (1000 points in total) | | | Actual data quantity (no repeat) | | | Number of involved cases |
|---|---|---|---|---|---|---|---|---|
| | | 10-cell | 15-cell | 20-cell | Total | 10-cell | 15-cell | |
| $u_g$ | 0.4499 | 733 | 267 | 0 | 442 | 267 | 146 | 36 |
| $u_l$ | 0.3902 | 666 | 318 | 16 | 495 | 282 | 197 | 35 |
| $\alpha$ | 0.5355 | 884 | 116 | 0 | 372 | 281 | 91 | 33 |

To explore whether this level of data similarity is sufficient for the training of DFNN model, some other database is also generated based on global similarity instead of local similarity. According to the values of global conditions (vapor and liquid injection rate, injection void fraction) in Case 35, different database can be built with different relationship with Case 35, as shown in Table 8. Measured by KDE similarity measure metric, their levels of physical feature similarity with Case 35 are much smaller than these of optimal training database shown in Table 7. The results denote that selecting optimal training database based on local similarity instead of global similarity can achieve higher level of data similarity and better predictive performance. Therefore, considering that $S_{opt,i} \geq S_{limit,i}$ and the good predictive performance shown in Step 2.4, the optimal training database selected in Table 8 has enough data similarity to predict the simulation errors of QoIs for Case 35.

Table 8. Summary of training database built based on global similarity.

| Backup training database # | Number of cases | Physical feature data similarity with Case 35 for QoIs ($S_{KDE}$) | | | Description of cases included |
|---|---|---|---|---|---|
| | | $u_g$ | $u_l$ | $\alpha$ | |
| 1 (a), (c), (e) | 18 | 0.2828 | 0.3758 | 0.4198 | (a) Same $u_g$ (0.347); |
| 2 (a), (c), (f) | 29 | 0.2708 | 0.3473 | 0.3978 | (b) Similar $u_g$ (0.293~0.347); |
| 3 (a), (c), (f) | 25 | 0.2864 | 0.3830 | 0.4291 | |
| 4 (a), (d), (f) | 33 | 0.2727 | 0.3548 | 0.4058 | (c) Same $u_l$ (1.087); |
| 5 (b), (c), (e) | 21 | 0.2833 | 0.3776 | 0.4240 | |
| 6 (b), (d), (e) | 31 | 0.2718 | 0.3512 | 0.4019 | (d) Similar $u_l$ (0.974~1.391); |
| 7 (b), (c), (f) | 26 | 0.2859 | 0.3835 | 0.4300 | (e) Similar $\alpha$ (0.17~0.25) |
| 8 (b), (d), (f) | 34 | 0.2726 | 0.3560 | 0.4072 | (f) Similar $\alpha$ (0.12~0.35) |
| 9 All cases | 41 | 0.2695 | 0.3488 | 0.3988 | |
| Case study in Step 2.4 ($S_{limit,i}$) | | 0.2722 | 0.2292 | 0.3217 | |

*4.4. Step 4: error prediction for coarse-mesh CFD*

In this step, the error prediction of local QoIs can be performed for Case 35 by using the optimal physical feature groups (Table 6) and the optimal training database (Table 7). By adding the DFNN predicted simulation errors, the modified coarse-mesh CFD simulation results are compared with the high-fidelity results as shown in Figure 11. Informed by previous low-fidelity and high-fidelity simulation results, the simulation errors of a new extrapolative case using coarse-mesh CFD can be well predicted using the proposed data-driven approach (Feature Similarity Measurement). Prediction errors are listed in Table 9.

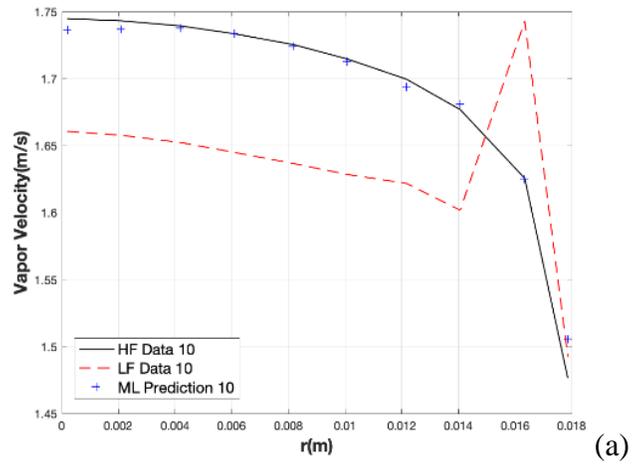

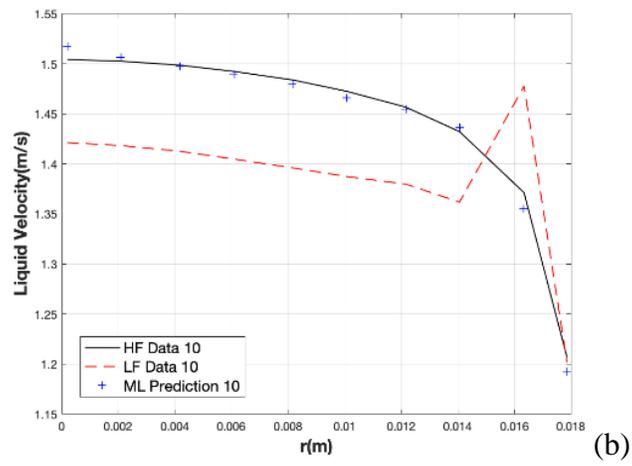

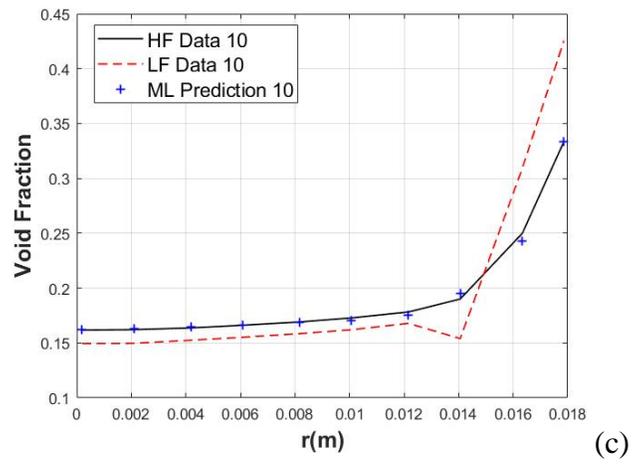

Figure 11. Comparison between original and DFNN-predicted coarse-mesh CFD simulation results with high-fidelity simulation results for Case 35.

Table 9. NRMSEs of DFNN predictions and low-fidelity simulations for Case 35.

| | $NRMSE_{u_g}$ | $NRMSE_{u_l}$ | $NRMSE_\alpha$ | Mesh configuration |
|---|---|---|---|---|
| DFNN prediction | 0.0057 | 0.0059 | 0.0158 | 10 cells |
| Low-fidelity simulation | 0.0499 | 0.0564 | 0.1934 | |

## 5. Discussions

The results of the case study show that FSM has the capability to estimate coarse-mesh CFD simulation errors. The CFD simulation with 10-cell configuration only has a total number of cells in the domain as 0.07 million while the 25-cell configuration validated as high-fidelity simulation has 0.63 million cells. Compared with the fine-mesh high-fidelity simulation, coarse-mesh simulation with FSM correction has a comparable accuracy and affordable computational cost. For the 10-cell coarse mesh CFD case, the total core-hours including simulation time and training cost are only ~6.6% of those of the high fidelity 25-cell configuration case.

Even the unphysical "peaks" near the wall in the 10-cell configuration are well captured and corrected by the well-trained DFNN model. The capability of capturing reginal patterns results from identifying 1-order and 2-order derivatives of QoIs as the physical features, because not only the characteristics at this point are captured, but also the connections of this point with its neighboring points. Another part of physical features defined as non-dimensional local physical parameters enable FSM the capability for extrapolative predictions. Similarity of local patterns between existing cases and the target case is measured by quantifying their similarity of local physical features. One hypothesis of FSM is, with higher level of similarity between training data and target data, the DFNN model can be better trained and has better predictive performance. A test about extrapolation of large liquid velocity was developed to evaluate this hypothesis: Case 37~42 are used as testing cases and Case 1~ 35 are used for training. The physical feature data similarity between these testing cases and the training database is measured using KDE method, and the evaluation metric for prediction accuracy is NRMSE. Only the optimal physical feature groups identified in the case study are considered in this test. As shown in Figure 12, NRMSE declines when the mean of KDE increases, which implies there is a positive relationship between data similarity and prediction accuracy.

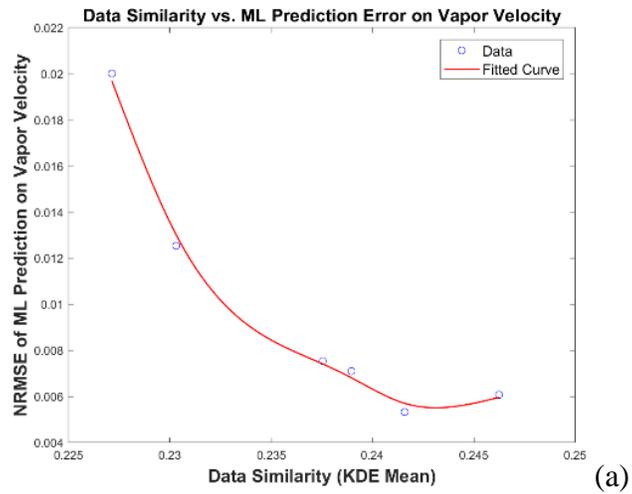 (a)

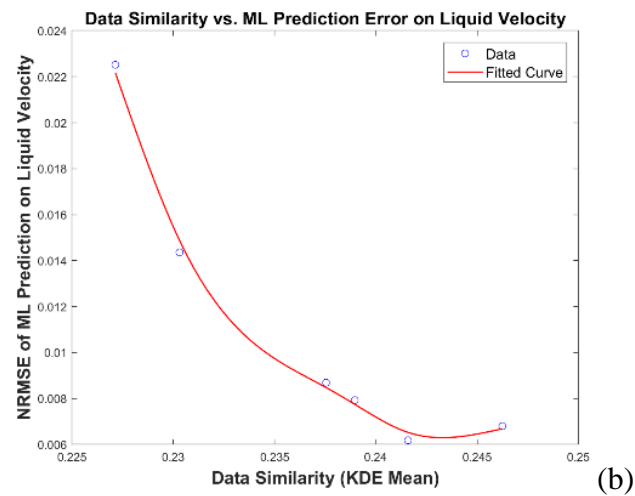 (b)

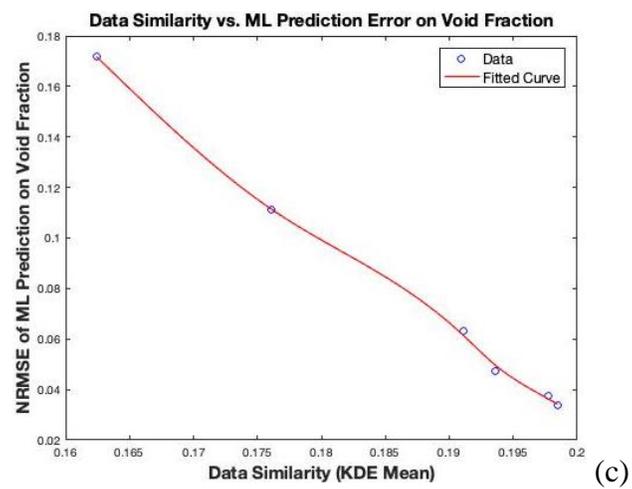 (c)

Figure 12. Relationship between data similarity and DFNN prediction accuracy.

## 6. Conclusions

This paper applied a physics-guided data-driven approach (Feature Similarity Measurement, FSM) to enable the computationally efficient CFD prediction. Including four steps as target analysis, feature identification, training database construction and error estimation, the proposed framework provides a data-driven guidance to improve the coarse-mesh CFD modeling and simulation capability with a substantially reduced computational cost. The modularized workflow was demonstrated based on a bubbly flow case study. By defining physics-guided local physical parameters and variable gradients as physical features, FSM exhibits the capability to capture and correct unphysical patterns in the coarse-mesh simulation, even for extrapolative conditions. The DFNN model well captured and corrected the unphysical "peaks" in the velocity and void fraction profiles near the wall in the coarse-mesh configuration, even for extrapolations.

With the application of advanced statistical algorithms and methods, the proposed workflow for computationally efficient CFD simulation also investigates how data affects the prediction accuracy of data-driven surrogate models. The importance of physical features on the error estimation of QoIs was evaluated, each QoI has its own "optimal" physical feature group, not all the pre-defined physical features are necessary for the error estimation of QoIs. Data similarity measurement is introduced for the construction of training database. According to the local similarity instead of global similarity, an optimal training database can be constructed with a higher level of data similarity, which only includes the relevant and sufficient data for the training of DFNN models. Using appropriate metrics for the evaluation of predictive accuracy and data similarity, suggestions are provided on how to choose the optimal physical feature group and optimal training database in order to make a balance between prediction accuracy and efficiency.

**Acknowledgement**

This work is supported by the U.S. Department of Energy, under Department of Energy Idaho Operations Office Contract DE-AC07-05ID14517. Accordingly, the U.S. Government retains a nonexclusive, royalty-free license to publish or reproduce the published form of this contribution, or allow others to do so, for U.S. Government purposes. The authors also would like to acknowledge the support from Baglietto CFD Research Group at MIT.

**Appendix—Discussion of Eulerian-Eulerian two-fluid model**

Numerical simulation of multiphase flows, especially multiphase computational fluid dynamics (M-CFD) approach, emerges as an effective tool for exploring the multiphase flow thanks to the increasing computer power and advanced algorithm. Due to the wide range of spatial and temporal scales in the industrial size system, it is virtually impossible to capture all the details of the flow field with the current available computational resources. Depending on the scales, three approaches are mostly used to simulate bubbly flows: the Eulerian-Eulerian (E-E) approach, the Eulerian-Lagrangian (E-L) approach and the Interface Tracking Methods (ITM) which is often coupled with direct numerical simulation (DNS) or large eddy simulation (LES) for turbulence modeling. These three approaches have their own advantages and disadvantages and their specific range of applicability. In the E-L approach, each bubble is separately tracked while the liquid phase is treated as a continuum. The interaction between the bubbles and liquid is accounted for through a source term in the momentum equations. In the ITM approach, individual bubble is tracked by resolving the interface which enables the investigation of the physical mechanism from a fundamental point of view. In the E-E approach, which is also referred to as the two-fluid model (Ishii and Mishima, 1984; Lahey and Drew, 2001; Podowski, 2009), both phases are treated as continuum fluids. The ensemble-averaged mass and momentum conservation equations are used to describe the time-dependent motion of both phases. At this level of E-E two-fluid models, bubbles lose their discrete identity, which enables the simulation of relatively large systems and makes E-E widely used in the industrial application compared to E-L and ITM approaches. Therefore, the E-E two-fluid model is used for the demonstration of the proposed framework.

For the E-E approach, the equations of the two-fluid model are derived by ensemble-averaging the local instantaneous equations. The governing mass conservation equations for adiabatic $N_p$ phases:

$$\frac{\partial}{\partial t}(\alpha_k \rho_k) + \nabla \cdot (\alpha_k \rho_k \overrightarrow{V_k}) = 0 \qquad (24)$$

The momentum equation is

$$\frac{\partial}{\partial t}(\alpha_k \rho_k \overrightarrow{V_k}) + \nabla \cdot (\alpha_k \rho_k \overrightarrow{V_k}\overrightarrow{V_k}) = -\alpha_k \nabla p_k + \nabla \cdot [\alpha_k (\tau_k + \tau_k^T)] + \alpha_k \rho_k g_k + M_k \qquad (25)$$

where $\alpha_k$ is the volume fraction of phase-$k$, $\rho_k$ is the density of phase-$k$, $\vec{V_k}$ is the velocity vector, $\tau_k$ is the shear stress tensor, $\tau_k^t$ is the turbulent shear stress tensor, $p_{ki}$ is the interfacial pressure and $M_k$ is the interfacial momentum transfer terms. Note that one should solve separate set of continuity and momentum equations for each phase along with the following condition: $\sum_{k=1}^{N} \alpha_k = 1$.

The momentum exchange between two phases is accounted for by the source term which can be expressed as a superposition of terms representing different physical mechanism, specifically

$$M_k = M_k^D + M_k^L + M_k^{VM} + M_k^W + M_k^{TD} \tag{26}$$

where the individual terms on the right hand side are: drag force (Ishii and Chawla, 1979; Tomiyama et al., 1998), lift force (Drew and Lahey, 1987; Tomiyama et al., 2002), virtual mass force (Drew and Lahey, 1987), wall lubrication force (Antal et al., 1991; Lubchenko et al., 2018; Tomiyama et al., 1995) and turbulent dispersion force (Lahey et al., 1993; Lopez de Bertodano, 1998; Podowski, 2008). Currently, it is generally agreed that the interfacial drag force is largely predominant over other distributions, and the lateral distribution of the bubble is tightly associated with the lift force. The accuracy of the closure relations for the interfacial forces significantly determines the prediction capability of the two-fluid model for the dispersed two-phase flow. In this paper, a set of those closures is adopted from Sugrue et al. (Sugrue et al., 2017) which is referred as the Bubbly and Moderate Void Fraction (BAMF) model and has been tested and validated for 12 cases from the Liu and Bankoff experimental datasets.

Along with the interfacial momentum closures, turbulent shear stress tensor, $\tau^t$, also need to be closed for the momentum equations. The most widely used model in the industries is the two-equation eddy viscosity model which is based on the Boussinesq hypothesis. It states that the anisotropy of the stress tensor is proportional to the mean velocity gradients:

$$\tau^t = 2\mu_t S_{ij} - \frac{2}{3}\rho k \delta_{ij} \tag{27}$$

These models transport the turbulent kinetic energy as well as the second variable, which is used to calculate the turbulent length scale. A typical choice is the dissipation rate of turbulent kinetic energy $\varepsilon$. In this case, turbulent eddy viscosity, $\mu_t$, is modeled as (Jones and Launder, 1972; Launder and Sharma, 1974)

$$\mu_t = C_\mu \rho \frac{k^2}{\varepsilon} \tag{28}$$

where $C_\mu$ is a model constant having a value of 0.09. Transport equations for the standard $k - \varepsilon$ model can be written as

$$\frac{\partial}{\partial t}(\rho k) + \nabla \cdot (\rho \vec{V} k) = \nabla \cdot \left[\left(\mu + \frac{\mu_t}{\sigma_k}\right)\nabla k\right] + P - \rho\varepsilon \tag{29}$$

$$\frac{\partial}{\partial t}(\rho \varepsilon) + \nabla \cdot (\rho \vec{V} \varepsilon) = \nabla \cdot \left[\left(\mu + \frac{\mu_t}{\sigma_\varepsilon}\right)\nabla \varepsilon\right] + C_{\varepsilon 1}\frac{\varepsilon}{k}P - C_{\varepsilon 2}\frac{\rho\varepsilon^2}{k} \tag{30}$$

where $\sigma_k$, $\sigma_\varepsilon$, $C_{\varepsilon 1}$, and $C_{\varepsilon 2}$ are model coefficients, and $P_k$ is a production term:

$$P = -(\rho \vec{V}' \vec{V}'):(\nabla \vec{V}) \tag{31}$$

The model coefficients are usually taken to be (Launder and Sharma, 1974):

$$C_{\varepsilon 1} = 1.44, C_{\varepsilon 2} = 1.92, \sigma_k = 1.0, \sigma_\varepsilon = 1.3 \tag{32}$$

For the two-phase turbulence model, the main assumption suggests that the Reynolds stress is negligible in the gas phases because the stress tensor is proportional to the phase density. Then the standard $k - \varepsilon$ equations are scaled proportionally to the volume fraction of liquid and expressed as followed where additional bubble-induced source terms $S_k$ and $S_\varepsilon$ are introduced

$$\frac{\partial}{\partial t}(\alpha_l \rho_l k_l) + \nabla \cdot (\alpha_l \rho_l k_l \vec{V_l}) = \nabla \cdot \left[\alpha_l\left(\left(\mu_l + \frac{\mu_l^T}{\sigma_{k_l}}\right)\nabla k_l\right)\right] + \alpha_l P - \alpha_l \rho_l \varepsilon_l + S_k \tag{33}$$

$$\frac{\partial}{\partial t}(\alpha_l \rho_l \varepsilon_l) + \nabla \cdot (\alpha_l \rho_l \varepsilon_l \vec{V_l})$$
$$= \nabla \cdot \left[\alpha_l\left(\left(\mu_l + \frac{\mu_l^T}{\sigma_{\varepsilon_l}}\right)\nabla \varepsilon_l\right)\right] + \alpha_l C_{\varepsilon_1}\frac{\varepsilon_l}{k_l}P - \alpha_l \rho_l C_{\varepsilon 2}\frac{\varepsilon_l}{k_l} + S_\varepsilon \tag{34}$$

where $\mu_l^T = \mu_t + \mu_{BIT}$ and $\mu_{BIT}$ are additional bubble-induced viscosity terms. In this paper, standard $k - \varepsilon$ equations are adopted for both liquid and gas phases and bubble-induced turbulence terms are deactivated for now to be consistent to the BAMF models (Sugrue et al., 2017).